\documentclass[prd,aps,superscriptaddress,nofootinbib,amsmath,amssymb,showpacs,9pt]{revtex4}
\usepackage{graphicx}
\usepackage{braket}
\usepackage{nicefrac}
\usepackage{float}%
\usepackage{amsmath}
\usepackage{hyperref}
\usepackage{multirow}
\usepackage{slashed}
 \usepackage{bbold} 

\usepackage[usenames, dvipsnames]{color}

\def\slashchar#1{\setbox0=\hbox{$#1$}
   \dimen0=\wd0 \setbox1=\hbox{/} \dimen1=\wd1
   \ifdim\dimen0>\dimen1 \rlap{\hbox to \dimen0{\hfil/\hfil}} #1
   \else  \rlap{\hbox to \dimen1{\hfil$#1$\hfil}} / \fi}

\begin{document}

\title{50 years of Neutrino Physics at Aligarh Muslim University}

\author{S. K. \surname{Singh}}
\affiliation{Department of Physics, Aligarh Muslim University, Aligarh-202 002, India}
\author{M. Sajjad \surname{Athar}}
\email{sajathar@gmail.com}
\affiliation{Department of Physics, Aligarh Muslim University, Aligarh-202 002, India}


\begin{abstract} 
An overview of the significant contributions made by 
 the Aligarh group in the field of neutrino physics has been provided. The group's work began with the study of 
 quasielastic neutrino scattering on deuterium, specifically the process $\nu_\mu + d \rightarrow \mu^- + p +p$, focusing 
 on moderate neutrino energies within the impulse approximation framework in order to analyze the early experiments at 
 Argonne National Lab (ANL) and Brookhaven National Lab (BNL) with deuterium filled bubble chambers. These studies were later extended to numerically 
 calculate (anti)neutrino interaction cross sections for the inclusive quasielastic scattering from nuclei using 
 local density approximation~(LDA), incorporating nucleon-nucleon correlations for moderate to heavy nuclear targets. The inelastic scattering cross sections for one pion production and the deep inelastic scattering~(DIS) from nuclei have been obtained using LDA with multi nucleon correlation effects.  
In the case of single pion production, the final state interaction of the pions with the residual nucleus has also been taken into account.
 In addition, the group has explored other inelastic processes of kaon and eta meson production, in the (anti)neutrino interactions with the nucleon targets. Furthermore, we have investigated electron and positron scattering off proton targets in the intermediate energy range, focusing on polarization observables and T-noninvariance. Our research has also delved into polarized electron scattering to explore the possibility of observing parity-violating asymmetry (PVA) in elastic and deep inelastic scattering, among other related topics. In the neutral current sector, we have also studied the quasielastic and inelastic 1$\pi^0$ production in order to explore $\nu(\bar\nu)$ magnetic moment effects by comparing the weak and electromagnetic induced reactions. The group has also contributed to the study of 
 atmospheric neutrino flux in the cosmic ray interactions, relevant for various sites of the neutrino oscillation experiments with atmospheric neutrinos.
\end{abstract}
\maketitle

\section{Introduction}
\label{intro}

The study of neutrino physics at Aligarh Muslim University~(AMU) started in 1974 and has continued for almost 50 years mainly in the field of neutrino
interactions with nucleons and nuclei and its applications to explore {\bf (i)} the structure of nucleons and nuclei, {\bf (ii)} to test the weak interaction theory, and 
{\bf (iii)} to understand the neutrino properties and their interactions. 
The early 1970's were exciting times in the neutrino physics when the era of high energy
(anti)neutrino experiments in the energy region of a few tens of GeV with accelerator neutrinos at CERN in Europe as well as at BNL and ANL in USA were reporting
results. The Standard Model~(SM) of Particle Physics proposed for leptons was extended to the hadron sector and was shown to be renormalisable. The Lorentz and isospin structure of weak hadronic currents predicted in the SM, which unified the 
electromagnetic and weak interactions were being tested in various processes of neutrino scattering from nucleons and nuclei. A new era in neutrino physics
had started. 

In the energy region of a few GeV of (anti)neutrinos, the processes of the elastic,  quasielastic, inelastic and deep inelastic 
scattering induced by the charged current (CC)  in the $\Delta S=0$ and $\Delta S=1$ sectors and by the neutral current (NC)  in the $\Delta S=0$ sector are important. The theoretical work has been done
on all these processes in the Department of Physics at AMU, which has gained international recognition and members of our working 
group are officially 
participating in the analysis of the neutrino experiments being done by the international collaboration of MINER$\nu$A and are involved in the development of the DUNE next generation experiment at the Fermi
National Accelerator Lab (FNAL) in USA. Some of our group members are also participating individually in the international collaboration of the 
ICARUS and SpinQuest experiments at the FNAL. In addition, we have also collaborated on the calculations of atmospheric neutrino flux 
with the group of scientists at the Institute for Cosmic Ray Research(ICRR) in the University of Tokyo, Japan
led by T. Kajita~(Nobel Prize winner in Physics-2015). The calculations have been done for the various experimental sites like INO in 
India, SuperKamiokande in Japan, IceCube in South Pole,  Pyhasalmi in Finland, etc. 
Our work has been extensively published in the international journals of 
repute like Physical Review, Journal of Physics, European Journal of Physics, Nuclear Physics, etc.

In the following, we give a brief account of the work done at AMU during the last 50 years on the processes of the quasielastic (QE) scattering, inelastic (IE) scattering, deep inelastic scattering (DIS) as well as the calculations of the  atmospheric neutrino flux and the weak interaction effects in the 
electron sector predicted by the SM.

\section{Quasielastic Scattering}
The early neutrino experiments on the quasielastic scattering done with the nuclear targets at ANL had reported anomalous results on the differential 
cross sections $\frac{d\sigma}{d Q^2}$ regarding the nuclear medium effects. In view of this, new experiments were done with the 
deuterium target in order to mitigate the nuclear medium effects. At AMU, the first work was done to study the nuclear effects in the 
neutrino-deuteron scattering, i.e.,
$\nu_\mu +d \to \mu^- + p +p$~\cite{Singh:1974df, Singh:1975zn}, which was an extension of earlier work~\cite{Singh:1971md}. The work was later extended to include 
the effect of meson exchange currents (MEC) in collaboration with the Nuclear Physics group at the University of Mainz~\cite{Singh:1986xh}. It was shown that the nuclear 
effects in deuteron reduce the cross section $\frac{d\sigma}{d Q^2}$ in the region of very low $Q^2$, where it could be $\sim 30\%$ at $Q^2\simeq 0$. The effect of MEC is to increase the cross sections by about 10\% at $Q^2\simeq 0$. The earlier work was used to analyze the first
neutrino experiments at ANL and BNL in order to determine the axial dipole mass $M_A$ occurring in the definitions of dipole 
parameterisation of axial form factor $g_A(Q^2)$, i.e.
\begin{equation}
 g_A(Q^2)=\frac{g_A(0)}{\Big(1+\frac{Q^2}{M_A^2} \Big)^2}
\end{equation}
after correcting for the deuteron effects.

The standard model of electroweak interactions predicted the NC interactions in the electron and neutrino sector with well determined 
Lorentz and isospin structure with no isoscalar axial vector currents, and vector currents given in terms of the electromagnetic current 
and the weak mixing angle $\theta_W$. In view of this, the NC disintegration of deuterium induced by $\nu(\bar\nu)$ 
i.e. $\nu (\bar\nu) + d \to \nu (\bar\nu) + n +p$ was investigated in detail in order to test the Standard Model and determine
the weak mixing angle. The other variants of current-current interactions in the NC sector with the scalar, tensor and pseudoscalar 
currents were also explored. Later some work was also done in the NC sector to test the presence of strangeness content of the nucleon 
through the process of neutrino-nucleus scattering~\cite{Singh:2001wv}.

In later years, the effect of nuclear medium on $M_A$ was studied in neutrino-nucleus scattering in collaboration with the theoretical 
physics group at the University of Valencia, Spain~\cite{Singh:1992dc, Singh:1993rg}. This work was then used to analyze the neutrino-nucleus cross sections 
in $^{12}C$ measured at Los Alamos and applied to study the experiments done with the atmospheric neutrinos in the energy region of a few hundreds of MeV. The theoretical work considered the inclusive (anti)neutrino inclusive reactions on $^{12}C$ and $^{40}Ar$ i.e. $\nu_\mu(\bar\nu_\mu) +^{12}C \to \mu^-(\mu^+) +X$ and $\nu_\mu(\bar\nu_\mu) +^{40}Ar \to \mu^-(\mu^+) +X$ . The nuclear medium effects due to the Fermi motion and nuclear binding were included in the local Fermi Gas model. In this model the effect of the long range nucleon nucleon correlations is calculated using the random phase approximation (RPA) in the ladder approximation. The long range nucleon-nucleon correlations are treated through $\pi$ and $\rho$ exchanges in the nucleon-nucleon potential
and the short range correlations are simulated through the use of Landau-Migdal parameter determined phenomenologically. It was 
shown that the effect due to the long range correlations and other nuclear effects are to reduce the cross sections by a factor of 40-60\%
in the range of a few hundreds of MeV of the incident (anti)neutrino energies relevant for the oscillation experiments with accelerator and atmospheric (anti)neutrinos. The formalism has also been applied in the low energy regime to study the response of the supernovae detectors,
which use water, hydrocarbon or lead nuclei as target material in the detector configuration and
also relevant for the SNS~\cite{SajjadAthar:2004yf} and KDAR neutrinos~\cite{Akbar:2017dih}, etc.

 The work on the quasielastic scattering was later extended to study the hyperon~($Y$) production by the antineutrinos from the nuclei like 
$^{12}C$, $^{16}O$, etc. and to study the nuclear medium effects as well as the final state interaction effects due to the $YN$ interaction
in the nucleus~\cite{Fatima:2018wsy, SajjadAthar:2022pjt}. The feasibility
of observing hyperon experimentally has been explored.  Moreover, the hyperon production
of $\Lambda$ and $\Sigma$ particles and their polarization facilitate the study of the second class currents and T-violation in weak interaction~\cite{Fatima:2018tzs}
through the study of asymmetries in the angular distributions of pions, which originate as a result of decays of $\Lambda$ 
and $\Sigma$ through $\Lambda (\Sigma) \to N \pi$ mode. Since hyperons have longer lifetimes due to their weak decays; the 
pions are produced outside the 
residual nucleus, therefore, avoiding the pion final state interactions with the residual nucleus, thus, reducing the theoretical uncertainties
in the calculations. The pions originating
    from the $\Lambda(\Sigma)$ decays are produced far from the $\Lambda(\Sigma)$ production point in the target nucleus due to longer lifetimes of 
    $\Lambda(\Sigma)$ and do not undergo the suppression due to the final state interactions. The pion production through $\Lambda(\Sigma)$ are therefore comparable
    to the pion production through $\Delta$ excitation in the low energy regions.
The work has been extended to quasielastic $\tau$ production including the $\tau$ polarization effects as this is important in detection 
of $\tau-$neutrinos ($\nu_\tau$)~\cite{Fatima:2020pvv, Fatima:2021fen}. 

To explore the possibility of finding useful limits, if any, on the neutrino magnetic moment
in the neutrino nucleus reactions induced by the electromagnetic current, the Aligarh group considered the 
elastic and inelastic transitions in some nuclei and found that there is a possibility of putting a 
better limit on the neutrino magnetic moment provided low energy
bolometric detectors are developed to be able to measure the very small recoil energies of 
nuclei in the neutrino-nucleus scattering~\cite{Singh:1995hv}.

\section{Inelastic scattering}
\subsection{One pion production}
The effect of nuclear medium in the quasielastic scattering of (anti)neutrinos from nuclei in the energy region of a few hundreds of MeV was found to 
be large and important in the analysis of neutrino oscillation experiments. After the phenomenon of neutrino oscillation was confirmed
in the atmospheric neutrinos, where the (anti)neutrino energies have a wide range covering the entire energy region in which the inelastic
reactions, especially the pion production induced by the charged and neutral currents, i.e. $\nu_\mu (\bar\nu_\mu) +N \to \mu^- (\mu^+) +N^\prime +\pi^\alpha;\;\alpha=\pm,0$
and $\nu (\bar\nu) +N \to \nu (\bar\nu) +N^\prime +\pi^\alpha$ are possible, the study of nuclear effects in the pion production in nuclei like $^{16}O$ 
became important as water target was used in the Kamiokande and IMB experiments, which reported the evidence for neutrino oscillations. Even though the weak production of pions
induced by the (anti)neutrinos from the nucleons has been studied for a long time~\cite{Athar:2020kqn, SajjadAthar:2022pjt}, the calculation of nuclear medium effects in the
case of nuclear targets was not done in the energy region of a few hundreds of MeV. Our group in Aligarh was one of the first groups to study these effects using the 
$\Delta-$dominance model in collaboration with the theoretical nuclear physics group from the University of Valencia, Spain~\cite{Singh:1998ha}. It was shown that the pion 
production cross sections are comparable to the quasielastic cross sections in the energy region of around 1 GeV and dominate for the (anti)neutrino energies $E_{\nu(\bar\nu)}>1$ GeV.
 The nuclear medium effects and the final state interaction of pion reduce the cross sections by a large amount of about 40\% around $E_{\nu(\bar\nu)}\simeq 1$ GeV
 but the ratio of the $\nu_e$ and $\nu_\mu$ induced cross sections, i.e., $\frac{\sigma(\nu_\mu)}{\sigma(\nu_e)}$, which was relevant for the neutrino oscillation 
 results was not affected. However, the fact that absolute cross sections are greatly affected by the nuclear medium effects started immense interest in 
 the study of pion production with (anti)neutrino scattering from the nuclear targets in the medium energy range~\cite{Athar:2020kqn, SajjadAthar:2022pjt}. Moreover, 
 the work on the weak pion production induced by the (anti)neutrinos in the few GeV energy region was 
 started by many groups going beyond the $\Delta-$dominance by including the effect of the non-resonance background terms as well as the higher resonances and our group 
 also contributed to this work. For details please see Ref.~\cite{SajjadAthar:2022pjt}. In the meantime, globally much interest was 
 generated in the coherent production of pions induced by the (anti)neutrinos from the nuclear targets due to very small cross sections inferred in some experiments
 as compared to the incoherent production in the region of intermediate energy of a  few GeV. A pioneering work was done by our group showing that suppression occurs 
 due to the nuclear medium effects and the final state interactions of pions with the residual nucleus~\cite{Singh:2006br}. The final state interaction was treated in Glauber
 model using a pion optical potential in the case of coherent production~\cite{Athar:2020kqn}. On the other hand, the final state interaction in the case of inelastic pion
 production was treated using a Monte Carlo simulation in which the pion suffers elastic scattering, pion charge exchange scattering and pion absorption while
 propagating inside the nucleus. In this approximation, a pion of a given
momentum and charge is moved along the z-direction with a
random impact parameter ${\bf b}$, with $|{\bf b}|<R$. Here $R$ is the
nuclear radius which is taken to be a point where nuclear density
$\rho(R)$ falls to ${10}^{-3}\rho_0$ with $\rho_0$ being the central
density. To start with, the pion is placed at a point $({\bf b},
z_{in})$, where $z_{in}=-\sqrt{R^2-|{\bf b}|^2}$ and then it is moved in
small steps $\delta l$ along the z-direction until it comes out of the
nucleus or interact. If $P(p_\pi,r,\lambda)$ is the probability per
unit length at the point $r$ of a pion of momentum ${\bf p}_\pi$ and
charge $\lambda$, then $P\delta l <<1$. A random number $x$ is generated
such that $x\in [0,1]$ and if $x > P\delta l$, then it is assumed
that pion has not interacted while traveling a distance
$\delta l$, however, if $x < P\delta l$ then the
pion has interacted and depending upon the weight factor of
each channel given by its cross section, it is decided that whether the
interaction was
 quasielastic, charge exchange reaction, pion production or pion absorption. For example, for the quasielastic scattering
\begin{equation*}
P_{N(\pi^\lambda,\pi^{\lambda^\prime})N^\prime}=\sigma_{N(\pi^\lambda,\pi^{\lambda^\prime})N^\prime}\times
\rho_N
\end{equation*}
where $N$ is a nucleon, $\rho_N$ is its density and $\sigma$ is the
elementary cross section for the reaction $\pi^\lambda +N \rightarrow
\pi^{\lambda^\prime} + N^\prime$ obtained from the phase shift
analysis. 

For a pion to be absorbed, $P$ is expressed in terms of the imaginary part
of the pion self energy $\Pi$
i.e. $P_{abs}=-\frac{Im\Pi_{abs}(p_\pi)}{p_\pi}$, where the self energy $\Pi$ is
related to the pion optical potential. 
 
  The nuclear medium effects in the incoherent production as well as in the 
 coherent production of pions is treated by modifying the $\Delta$ properties in the nuclear medium by changing its mass $M_\Delta$ and width $\Gamma$ given by
 \begin{eqnarray}
  M_\Delta &\to& M_\Delta +Re \Sigma_\Delta,\\
  -Im \Sigma_\Delta &=& C_Q\;\Big(\frac{\rho}{\rho_0}\Big)^\alpha+C_{A_2}\;\Big(\frac{\rho}{\rho_0}\Big)^\beta+C_{A_3}\;\Big(\frac{\rho}{\rho_0}\Big)^\gamma\\
  \frac{\Gamma}{2}&=&\frac{\Gamma}{2}-Im\Sigma_{\Delta},
 \end{eqnarray}
where $Re \Sigma_\Delta$, $C_Q$, $C_{A_2}$, $C_{A_3}$ and $\alpha$, $\beta$, $\gamma$ are determined phenomenologically~\cite{Singh:1998ha}. The cross section for one pion production from the nuclear targets in the presence of the nuclear medium effects and the final state interactions of pions were compared with the available experimental results.

\subsection{Neutral current induced $\pi^0$ production and neutrino magnetic moment}
 To look for putting a better limit on the neutrino magnetic moment ($\mu_\nu$), in the neutral current induced $\pi^0$ production from the nuclear 
 targets in the near future neutrino oscillation experiments, the Aligarh group studied, the weak $\pi^0$ production induced by 
 the neutral current in the standard model and compared the cross sections obtained with the electromagnetic induced  
 $\pi^0$ production with the present limits on $\mu_\nu$. It was found that with the present limits of $\mu_\nu$, the  
 $\pi^0$ production cross sections are quite smaller then the weak neutral current induced cross sections, and it 
 is difficult to put a better limit on $\mu_\nu$ using these processes~\cite{Athar:2008bv}.
\subsection{Two pion production}
The two pion production induced by the neutrinos i.e.
\begin{equation}
 \nu_l+N \to l^- +N^\prime +\pi^\alpha+\pi^\beta;\;\;\alpha,\beta=+,-,0
\end{equation}
is studied in the threshold region, in which the contribution comes from the nonresonance terms and the resonance terms~\cite{Hernandez:2007ej}. The nonresonance terms
are treated in an effective Lagrangian given by the SU(2) nonlinear $\sigma$ model while the contribution from the Roper resonance $N^\ast(1440)$ is considered,
which couples strongly to the two pion decay modes, i.e., $N^\ast(1440) \to N+\pi^\alpha+\pi^\beta$. The vector form factors for 
$\gamma N \to N^\ast$ transitions are determined from $e N \to e N^\ast$ transition while the axial vector part is determined using partially conserved axial-vector current(PCAC) and
the experimental strength of $N^\ast \to N\pi$ decay. It is found that at lower energies the contribution from the Roper resonance dominates
but at higher energies i.e., beyond $E_\nu>0.7$ GeV, the non-resonance background terms dominate.

 \subsection{$\eta$ and $K$ meson production} 
 The $\eta$ meson production occurs via $\Delta S=0$ charged and neutral currents, i.e., $\nu_\mu +n  \to \mu^-+p + \eta$, $\bar\nu_\mu +p  \to \mu^++n + \eta$
 and $\nu+p(n)  \to \nu+p(n) + \eta$, while the $K$ mesons can be produced by $\Delta S=0$ currents through the associated production induced by charged and neutral 
 currents, i.e. $\nu_\mu +n  \to \mu^-+\Lambda + K^+$, $\bar\nu_\mu +n  \to \mu^++\Lambda + K^-$ and $\nu_\mu(\bar\nu_\mu) +p  \to \nu_\mu(\bar\nu_\mu)+\Lambda + K^+$,
 $\nu_\mu(\bar\nu_\mu) +n  \to \nu_\mu(\bar\nu_\mu)+\Lambda + K^0$ but only through CC in the $\Delta S=1$ sector via. the reactions 
  $\nu_\mu(\bar\nu_\mu) +p  \to \mu^-(\mu^+)+p + K^+(K^0)$,   $\nu_\mu(\bar\nu_\mu) +n  \to \mu^-(\mu^+)+n + K^+(\bar K^0)$, 
    $\nu_\mu +n  \to \mu^-+p + K^0$ and  $\bar\nu_\mu +p  \to \mu^++n + \bar K^0$~\cite{SajjadAthar:2022pjt}. We have calculated the $\eta$ production from nucleons and found that it is 
    dominated by the $S_{11}(1535)$ resonance~\cite{Fatima:2022nfn}. The results have been compared with preliminary results from the experiment~\cite{Fatima:2023fez}. 
    
    The $K^+$, $K^0(K^+)$ production 
    from the neutrons(protons) induced by the neutrinos and $K^-$, $\bar K^0(K^-)$ production from the protons (neutrons) induced by the antineutrinos in the $\Delta S=1$ sector have been studied. Further details are given in Ref.~\cite{Athar:2020kqn, SajjadAthar:2022pjt}. The results from our group have already been incorporated in the neutrino generator GENIE(Generates Events for Neutrino Interaction Experiments), which is being used in the 
    various neutrino oscillation experiments. We are still working on the associated production induced by the weak currents. Towards this goal, we have already 
    produced the satisfactory results theoretically for the associated photoproduction i.e. $\gamma + p \rightarrow \Lambda + K^+$ and  $\gamma + n \rightarrow \Lambda + K^0$ in order to understand the importance of the role of various couplings
    in the vector sector.
    
    \subsection{Weak interaction studies with electron beams}
    The SM reproduces the V-A interaction in the charged current induced processes and predicts new interaction of neutral current in the electron sector. In view of the very
    high intensity and high energy electron beams available at the electron accelerators like SLAC, JLab and MAINZ, the weak interaction studies
    are performed with electron 
    beams in CC as well as in the NC sectors. In the CC sector, the quasielastic production of nucleons and delta resonances induced by $\Delta S=0$ transitions 
    like $e^-+p\to n+\nu_e$ and $e^- + p\to \Delta^0 + \nu_e$ and the hyperon production induced by the $\Delta S=1$ transitions like $e^-+p \to \Lambda (\Sigma)+\nu_e$
    were studied. The $\Delta$ production as well as the $\Lambda(\Sigma)$ production are more feasible as they can be observed through the pions, which 
    are produced when $\Delta$ and $\Lambda(\Sigma)$ decay through the $N\pi$ channels by the strong and weak interactions, respectively.   We have studied the weak production of $\Delta$, $\Lambda$, and $\Sigma$ baryons  induced 
    by electrons from nucleons~\cite{Akbar:2017qsf}. Recently, the work has been extended to study the feasibility of determining the sub-dominant $N-\Delta$ transition form factors
    in the weak electroproduction of $\Delta$~\cite{Fatima:2024hlu}. 
    
    In the NC sector, the weak interaction effects arise when the parity violating scattering amplitude through the $Z$ exchange, which interferes with the parity conserving 
    electromagnetic scattering amplitude through the $\gamma$ exchange and leads to the parity violating effects in the scattering of the polarized 
    electrons from the nucleons and nuclei. This process had been the signature process for confirming the existence of NC in the polarized electron scattering earlier at SLAC
    and have been confirmed at MAINZ, MIT and JLab electron accelerators.
    We have calculated these parity violating asymmetries in the elastic scattering of the polarized electrons from the deuteron~\cite{Murthy:1983wt, Arenhovel:2000if} and in the deep inelastic 
    scattering of the polarized electrons from the nuclei with the nuclear medium effects~\cite{Haider:2014iia}.
   
    \section{Deep Inelastic Scattering}
    Aligarh group has also studied the charged and neutral current deep inelastic scattering processes induced by both
the charged leptons and the (anti)neutrinos off the free nucleons as well as from the different nuclear targets. The group has published theoretical results 
for the nucleon structure functions, the  nuclear structure functions, the differential scattering cross sections, the total cross sections and 
the degree of polarization of the massive charged leptons.

In the electromagnetic sector, only two structure functions $F_{1N}(x,Q^2)$ and  $F_{2N}(x,Q^2)$ contribute
to the DIS cross section~\cite{SajjadAthar:2009cr}, while in the weak sector there is additional contribution from the parity violating 
structure function  $F_{3N}(x,Q^2)$ in the massless limit of charged leptons~\cite{Haider:2011qs, Zaidi:2024obq}. The group has evaluated these nucleon structure functions by performing the
higher order perturbative evolution of the parton densities up to the next-to-next-to-the-leading order  (NNLO) using the 
different nucleonic parton distribution functions~(PDFs) parameterizations like CTEQ6.6, CT14, MSTW, MMHT, etc., as well as by incorporating the 
non-perturbative corrections such as the target mass corrections (TMC) and the higher twist (HT) corrections. In the case of massive 
charged lepton production, additional structure functions $F_{4N}(x,Q^2)$ and $F_{5N}(x,Q^2)$ also contribute~\cite{Ansari:2020xne}.  The
target mass corrected structure functions are obtained using the operator product expansion approach while HT (twist-4) corrections are incorporated 
following the renormalon approach. 
It has been found that the nucleon structure functions $F_{2N}(x,Q^2)$ and $2 x F_{1N}(x,Q^2)$ get modified at high 
$x$ and low $Q^2$ due to the inclusion of HT effect when evaluated at the next-to-the-leading order~(NLO). However, for the low $x$ 
region, the impact of HT effect in $2 x F_{1N}(x,Q^2)$ is found to be more pronounced than in the case of
$F_{2N}(x,Q^2)$. The HT effect decreases with the increase in $Q^2$. $F_{1N}(x,Q^2)$ has also been evaluated 
independently beyond the leading order (LO) by taking the non-zero contribution from the longitudinal structure
function $F_{LN}(x,Q^2)$. This study has been helpful in understanding the behavior of the Callan-Gross relation
in the region of low and moderate $Q^2$. $F_{4N}(x,Q^2)$ is zero at the leading order but has non-negligible contribution
when higher order perturbative effects are taken into account.

When the scattering takes place with a nucleon bound inside a nuclear target, due to the presence of 
nuclear medium effects such as  binding energy, Fermi motion, multi-nucleon correlation effects, mesonic contributions, and shadowing 
and antishadowing contributions, the structure functions get modified. 
The Aligarh group has incorporated nuclear medium effects using a microscopic field theoretical approach. The effect of Fermi motion, binding energy 
and nucleon correlations are included through the relativistic nucleon spectral function, which is obtained by using 
the Lehmann's representation for the relativistic nucleon propagator, and the technique of nuclear many body theory is then used to calculate 
the dressed nucleon propagator in an interacting Fermi sea in the nuclear matter. To obtain the results for a finite nucleus, 
LDA is then applied. The mesonic contribution has been incorporated by using many-body field theoretical approach similar 
to the case of bound nucleons. Furthermore, the shadowing effect dominates in the region of low $x$, where the  
intermediate vector bosons create quark-antiquark pairs that interact with the partons. The multiple scattering of quarks causes the destructive
interference of amplitudes that leads to the phenomenon of shadowing. The Aligarh group has also used their theoretical model to study the nonisoscalarity correction for nuclei having $(A-Z)~~>>Z~~$, without using any
phenomenological prescription. 
The numerical results obtained for the deuterium, helium, carbon, oxygen, iron and lead targets are being used by the MINER$\nu$A collaboration in their analyses.
This model has also been used in the extraction of the weak mixing angle sin$^2\theta_W$
using Paschos-Wolfenstein (PW) relation~\cite{Haider:2012ic}. Efforts by this group has been made to study and understand the nuclear medium effects on the parity 
violating asymmetry in the scattering of the polarized electrons from the nuclear targets like $^{12}C$, $^{56}$Fe and $^{208}$Pb.

Recently this group has studied the effects of perturbative and nonperturbative QCD corrections to the Quark-Parton Model(QPM) in the evaluation 
of polarized 
nucleon structure functions by performing the calculations up to NNLO using LSS06 PPDFs in the 3-flavor MSbar scheme. 
 The nonperturbative effects of target mass correction and the twist-3 correction have also been incorporated.
The results are presented for the polarized proton and neutron structure functions, $g_{1p,1n}(x,Q^2)$ and $g_{2p,2n}(x,Q^2)$, the nucleon asymmetries 
$A_{1p,1n}(x,Q^2)$, $A_{2p,2n}(x,Q^2)$ as well as the Ellis-Jaffe, Bjorken, Burkhardt-Cottingham and Gottfried sum rules~\cite{Zaidi:2024obq}.


\section{Atmospheric Neutrino Flux}

The Aligarh group, in collaboration with the scientists lead by M. Honda and T. Kajita from the Institute for Cosmic Ray 
Research in Japan, has investigated the atmospheric neutrino flux at several experimental sites, including the SuperKamiokande,
the South Pole, Pyhasalmi, and the India-based Neutrino Observatory (INO)~\cite{SajjadAthar:2012dji, Honda:2015fha}. Their 
study utilized the Atmospheric Muon Neutrino Calculation (ATMNC) model, which simulates cosmic ray propagation through 
the atmosphere using JAM-a code incorporated in the Particle and Heavy-Ion Transport Code System (PHITS)
~\cite{SajjadAthar:2012dji, Honda:2015fha}.  

Despite employing the same primary flux and the interaction models for all the experimental sites, the team observed variations
in the calculated atmospheric neutrino flux due to the influence of the geomagnetic field. The geomagnetic field impacts the cosmic 
rays both inside and outside the atmosphere by filtering the low-energy cosmic rays and deflecting charged particles within
the atmosphere. These effects are primarily controlled by the horizontal component of the geomagnetic field.

The study revealed significant differences in the nature of the atmospheric neutrino flux, particularly at
low and intermediate energies, depending on the location within the geomagnetic field. The strength of
the horizontal geomagnetic component serves as a reliable indicator of deviations in the fluxes observed 
at different sites. Notably, the zenith angle dependence of the flux is critical in the neutrino oscillation
analyses, with larger discrepancies observed at lower neutrino energies, up to a few GeV. These findings 
are highly relevant for the interpretation of the atmospheric neutrino experiments proposed at these locations~\cite{SajjadAthar:2012dji, Honda:2015fha}.  

Additionally, the group assessed the uncertainties in the atmospheric neutrino flux calculations arising from the 
variations in the hadronic interaction model. They proposed a method to mitigate these uncertainties by 
using accurately measured atmospheric muon flux. By treating differences between the modeled and actual
hadronic interactions as a variation, the team developed a quantitative approach to estimate errors in the 
atmospheric neutrino flux calculations. This method relies on reconstruction residuals from the atmospheric
muon flux observations obtained through precision experiments~\cite{Honda:2019ymh}.

It was found that the correlation between the neutrino flux calculation errors and the muon flux reconstruction
residuals varies significantly based on the muon observation site, particularly for the low-energy neutrinos.
The study was conducted at several locations, including Kamioka (at sea level), 2770 meters above the sea level, Hanle
in India (4500 meters a.s.l.), and at balloon altitudes ($\sim$ 32 km). The researchers also evaluated how effectively 
atmospheric muon data can reduce the error margin in the atmospheric neutrino flux calculations.
    
    \section{Journey so far and the Future Prospects}
    Our work in the neutrino physics and the weak interaction physics in general has gained national and international 
    recognition and appreciation~\cite{Alvarez-Ruso:2020ezu, Ruso:2022qes, NuSTEC:2017hzk, SajjadAthar:2021prg, SajjadAthar:2020nvy}. We have contributed in most  of the national level meetings on the nuclear and particle physics and represent 
    in the Scientific Program Committee (SPC) of the only neutrino experiment INO since its inception. At the international level, we have been associated with the 
    MINER$\nu$A collaboration at Fermilab since the beginning of the experiment in 2006 and are now official members of the collaboration. At MINERvA, the research students from our group played pivotal roles across various cutting-edge initiatives. Some of them were involved in collecting data for antineutrino-nucleus scattering in the medium energy mode($<E_{\nu_\mu}> \sim 6$GeV). They contributed significantly to the infrastructure development for the Data Preservation project and led the Data Production team with exceptional efficiency. Their work encompassed advanced tasks such as vertex finding in neutrino-nucleus interactions, where they conducted a thorough comparison of model architectures. They also engaged in rigorous 1D and 2D antineutrino-nucleus CC inclusive and DIS analyses, demonstrating expertise in handling complex datasets. Additionally, they spearheaded the production of machine learning tuples, a critical component of the Data Preservation effort, ensuring the longevity and accessibility of MINERvA's invaluable data.
    Recently, our group has become 
    a member of the DUNE collaboration also at Fermilab and is actively involved in studying various $\nu$-processes, which will be studied by this collaboration.
    The members of our groups are interested and involved in the future experiments being planned by the ICARUS collaboration to study the neutrino oscillation and the SpinQuest collaboration at the Fermilab, 
    to study ``spin'' of the nucleon. The group has published more 
    than 120 research papers in peer reviewed journals and the papers listed in the references give only a representative 
    of papers published on various topics. The group has published several long and 
    comprehensive reviews~\cite{SajjadAthar:2022pjt,SajjadAthar:2020nvy,Fatima:2018wsy} and
    a text book on ``The Physics of Neutrino Interactions'' of about 950 pages published by the Cambridge University Press~\cite{Athar:2020kqn} 
    and edited 4 volumes of Conference proceedings.
    
    The group at Aligarh Muslim University is actively involved in planning various activities of the neutrino physics both theoretically and experimentally at the 
    International level as one of senior member is either serving or has served in the executive committee of the
    NUSTEC (Neutrino Scattering Theoretical and Experimental Collaboration), the IUPAP (International Union of Pure and Applied Physics) in its neutrino section.
    We have been actively associated with the Snowmass projects in carrying out future planning in the field of pion production
    and deep inelastic scattering and hope to collaborate in their future activities in neutrino physics. This group is also involved in bringing a white paper on 
     Long Range Plan for Nuclear Physics in Europe on behalf of the Nuclear Physics European Collaboration Committee.
    
    The group has also been associated with the NuInt series of international conference in the neutrino-nucleus interactions since its beginning in 2001
    and hosted the 10$^{th}$ meeting in 2011 in India and has been a regular contributor to the organization and participation in these meetings. In addition, the group
    members have also been associated with the scientific program committees of the NUFACT series of neutrino meetings since 2006.

    To summarize, our work on the various processes induced by the (anti) neutrinos on nucleons and nuclei has significant impact on neutrino physics and has made important contributions towards understanding the nuclear medium effects on the neutrino scattering cross sections in the energy region of few GeV relevant for the present neutrino oscillation experiments. To give a few examples:
    \begin{itemize}
\item 
The effect of using deuteron target was shown to be very important in the low $Q^2$ region of the quasielastic scattering and was used to analyse the early experiments at the Argonne and Brookhaven Labs. The work was extended by other groups later to study the deuteron effects in the pion production process.

\item The effect of the nuclear medium on the (anti)neutrino-nucleus scattering was shown to be very important for the quasi elastic and inelastic process of pion production in the region of intermediate energy  relevant for the present neutrino oscillation experiments. The work and its methodology has been extended by other groups to calculate the higher order correction to the nuclear medium effects. 
These calculations have been adopted in the most often used neutrino generator GENIE for simulating neutrino events.

\item In the region of the very high energy of (anti)neutrinos where the process of the deep inelastic Scattering is important. Our calculations of nuclear medium effects has been very useful for the neutrino physics community. Our work on the resonance excitations in the (anti)neutrino scattering  has contributed to the understanding of the shallow inelastic scattering region and its connection with the deep inelastic region.

\item Our calculations of the eta, kaon and hyperon production was the first calculation of these processes in the weak sector and some of the work has been incorporated in the neutrino generator GENIE and is being used in the analysis of the neutrino oscillation experiments. Additionally, our research on deep inelastic scattering cross sections for various nuclear targets has been instrumental in advancing the field and is actively utilized by the MINERvA collaboration to enhance their experimental analyses.

\item The calculations of the atmospheric neutrino flux conducted by our group member have improved upon the earlier work by Honda et al. These calculations have also extended the study of atmospheric neutrino fluxes to polar and tropical regions, corresponding to the sites of current and future neutrino oscillation experiments using atmospheric neutrinos. This work has significantly contributed to a better understanding of atmospheric neutrino fluxes, particularly in the low and intermediate energy regions.
\end{itemize}

 We see a very bright future for research activities in the neutrino physics in the years to come.


\begin{thebibliography}{100}
\bibitem{Singh:1974df}
S.~K.~Singh,
Phys. Rev. D \textbf{10}, 988-992 (1974),
doi:10.1103/PhysRevD.10.988.


\bibitem{Singh:1975zn}
S.~K.~Singh,
Phys. Rev. D \textbf{11}, 2602-2605 (1975),
doi:10.1103/PhysRevD.11.2602.


\bibitem{Singh:1971md}
S.~K.~Singh,
Nucl. Phys. B \textbf{36}, 419-435 (1972),
doi:10.1016/0550-3213(72)90227-1.


\bibitem{Singh:1986xh}
S.~K.~Singh and H.~Arenhovel,
Z. Phys. A \textbf{324}, 347-354 (1986),
doi:10.1007/BF01294589.


\bibitem{Singh:2001wv}
S.~K.~Singh and M.~Sajjad Athar,
J. Phys. G \textbf{27}, 883-892 (2001),
doi:10.1088/0954-3899/27/4/312.


\bibitem{Singh:1992dc}
S.~K.~Singh and E.~Oset,
Nucl. Phys. A \textbf{542}, 587-615 (1992),
doi:10.1016/0375-9474(92)90259-M.


\bibitem{Singh:1993rg}
S.~K.~Singh and E.~Oset,
Phys. Rev. C \textbf{48}, 1246-1258 (1993),
doi:10.1103/PhysRevC.48.1246.


\bibitem{SajjadAthar:2004yf}
M.~Sajjad Athar and S.~K.~Singh,
Phys. Lett. B \textbf{591}, 69-75 (2004),
doi:10.1016/j.physletb.2004.04.025.


\bibitem{Akbar:2017dih}
F.~Akbar, M.~Sajjad Athar and S.~K.~Singh,
J. Phys. G \textbf{44}, no.12, 125108 (2017),
doi:10.1088/1361-6471/aa9125.


\bibitem{Fatima:2018wsy}
A.~Fatima, M.~Sajjad~Athar and S.~K.~Singh,
Front. in Phys. \textbf{7}, 13 (2019),
doi:10.3389/fphy.2019.00013.


\bibitem{SajjadAthar:2022pjt}
M.~Sajjad Athar, A.~Fatima and S.~K.~Singh,
Prog. Part. Nucl. Phys. \textbf{129}, 104019 (2023),
doi:10.1016/j.ppnp.2022.104019.

\bibitem{Fatima:2018tzs}
A.~Fatima, M.~Sajjad Athar and S.~K.~Singh,
Phys. Rev. D \textbf{98}, no.3, 033005 (2018)
doi:10.1103/PhysRevD.98.033005.

\bibitem{Fatima:2020pvv}
A.~Fatima, M.~Sajjad Athar and S.~K.~Singh,
Phys. Rev. D \textbf{102}, no.11, 113009 (2020),
doi:10.1103/PhysRevD.102.113009.



\bibitem{Fatima:2021fen}
A.~Fatima, M.~Sajjad~Athar and S.~K.~Singh,
[arXiv:2111.13021 [hep-ph]].


\bibitem{Singh:1995hv}
S.~K.~Singh and M. Sajjad Athar,
Phys. Rev. C \textbf{52}, 2203-2209 (1995),
doi:10.1103/PhysRevC.52.2203.


\bibitem{Athar:2020kqn}
M.~Sajjad~Athar and S.~K.~Singh,
Cambridge University Press, 2020,
ISBN 978-1-108-77383-6, 978-1-108-48906-5,
doi:10.1017/9781108489065.


\bibitem{Singh:1998ha}
S.~K.~Singh, M.~J.~Vicente-Vacas and E.~Oset,
Phys. Lett. B \textbf{416}, 23-28 (1998)
[erratum: Phys. Lett. B \textbf{423}, 428 (1998)],
doi:10.1016/S0370-2693(97)01325-7.


\bibitem{Singh:2006br}
S.~K.~Singh, M.~Sajjad Athar and S.~Ahmad,
Phys. Rev. Lett. \textbf{96}, 241801 (2006),
doi:10.1103/PhysRevLett.96.241801.


\bibitem{Athar:2008bv}
M.~Sajjad~Athar, S.~Chauhan and S.~K.~Singh,
Phys. Rev. D \textbf{78}, 037301 (2008),
doi:10.1103/PhysRevD.78.037301.



\bibitem{Hernandez:2007ej}
E.~Hernandez, J.~Nieves, S.~K.~Singh, M.~Valverde and M.~J.~Vicente Vacas,
Phys. Rev. D \textbf{77}, 053009 (2008),
doi:10.1103/PhysRevD.77.053009.


\bibitem{Fatima:2022nfn}
A.~Fatima, M.~Sajjad Athar and S.~K.~Singh,
Phys. Rev. D \textbf{107}, no.3, 033002 (2023),
doi:10.1103/PhysRevD.107.033002.

\bibitem{Fatima:2023fez}
A.~Fatima, M.~Sajjad Athar and S.~K.~Singh,
Phys. Rev. D \textbf{108}, no.5, 053009 (2023)
doi:10.1103/PhysRevD.108.053009.

\bibitem{Akbar:2017qsf}
F.~Akbar, M.~Sajjad Athar, A.~Fatima and S.~K.~Singh,
Eur. Phys. J. A \textbf{53}, no.7, 154 (2017)
doi:10.1140/epja/i2017-12340-4.

\bibitem{Fatima:2024hlu}
A.~Fatima, M.~Sajjad Athar and S.~K.~Singh,
Phys. Rev. D \textbf{110}, no.5, 053009 (2024),
doi:10.1103/PhysRevD.110.053009.

\bibitem{Murthy:1983wt}
M.~V.~N.~Murthy, G.~Ramachandran and S.~K.~Singh,
Pramana \textbf{20}, 221-244 (1983),
doi:10.1007/BF02846216.
 

\bibitem{Arenhovel:2000if}
H.~Arenhovel and S.~K.~Singh,
Eur. Phys. J. A \textbf{10}, 183-205 (2001),
doi:10.1007/s100500170131.


\bibitem{Haider:2014iia}
H.~Haider, M.~Sajjad Athar, S.~K.~Singh and I.~Ruiz Simo,
Nucl. Phys. A \textbf{940}, 138-157 (2015),
doi:10.1016/j.nuclphysa.2015.04.001.

\bibitem{SajjadAthar:2009cr}
M.~Sajjad Athar, I.~Ruiz Simo and M.~J.~Vicente Vacas,
Nucl. Phys. A \textbf{857}, 29-41 (2011),
doi:10.1016/j.nuclphysa.2011.03.008.


\bibitem{Haider:2011qs}
H.~Haider, I.~R.~Simo, M.~Sajjad~Athar and M.~J.~V.~Vacas,
Phys. Rev. C \textbf{84}, 054610 (2011),
doi:10.1103/PhysRevC.84.054610.


\bibitem{Zaidi:2024obq}
F.~Zaidi, M.~Sajjad Athar and S.~K.~Singh,
[arXiv:2407.15410 [hep-ph]].


\bibitem{Ansari:2020xne}
V.~Ansari, M.~Sajjad Athar, H.~Haider, S.~K.~Singh and F.~Zaidi,
Phys. Rev. D \textbf{102}, no.11, 113007 (2020),
doi:10.1103/PhysRevD.102.113007.

\bibitem{Haider:2012ic}
H.~Haider, I.~R.~Simo and M.~Sajjad~Athar,
Phys. Rev. C \textbf{87}, no.3, 035502 (2013),
doi:10.1103/PhysRevC.87.035502.

\bibitem{SajjadAthar:2012dji}
M.~Sajjad Athar, M.~Honda, T.~Kajita, K.~Kasahara and S.~Midorikawa,
Phys. Lett. B \textbf{718}, 1375-1380 (2013),
doi:10.1016/j.physletb.2012.12.016.

\bibitem{Honda:2015fha}
M.~Honda, M.~Sajjad Athar, T.~Kajita, K.~Kasahara and S.~Midorikawa,
Phys. Rev. D \textbf{92}, no.2, 023004 (2015),
doi:10.1103/PhysRevD.92.023004.

\bibitem{Honda:2019ymh}
M.~Honda, M.~Sajjad Athar, T.~Kajita, K.~Kasahara and S.~Midorikawa,
Phys. Rev. D \textbf{100}, no.12, 123022 (2019),
doi:10.1103/PhysRevD.100.123022.

\bibitem{Alvarez-Ruso:2020ezu}
L.~Alvarez-Ruso, A.~M.~Ankowski, M.~Sajjad Athar \textit{et al.}
``Snowmass 2021 LoI: Neutrino-induced Shallow- and Deep-Inelastic Scattering,''
[arXiv:2009.04285 [hep-ex]].
 

\bibitem{Ruso:2022qes}
L.~A.~Ruso, A.~M.~Ankowski, .., M.~Sajjad Athar  \textit{et al.}
``Snowmass 2021 LoI: Theoretical tools for neutrino scattering: interplay between lattice QCD, EFTs, nuclear physics, phenomenology, and neutrino event generators,''
[arXiv:2203.09030 [hep-ph]]. J. of Phys. G (in press).


\bibitem{NuSTEC:2017hzk}
L.~Alvarez-Ruso, M.~Sajjad Athar \textit{et al.} [NuSTEC],
Prog. Part. Nucl. Phys. \textbf{100}, 1-68 (2018)
doi:10.1016/j.ppnp.2018.01.006.


\bibitem{SajjadAthar:2021prg}
M.~Sajjad Athar, S.~W.~Barwick, T.~Brunner, J.~Cao, M.~Danilov, K.~Inoue, T.~Kajita, M.~Kowalski, M.~Lindner and K.~R.~Long, \textit{et al.}
Prog. Part. Nucl. Phys. \textbf{124}, 103947 (2022)
doi:10.1016/j.ppnp.2022.103947.


\bibitem{SajjadAthar:2020nvy}
M.~Sajjad Athar and J.~G.~Morf\'\i{}n,
J. Phys. G \textbf{48}, no.3, 034001 (2021)
doi:10.1088/1361-6471/abbb11.
\end{thebibliography}
\end{document}